Long Paper

# A Study of an Agile Methodology with Scrum Approach to the Filipino Company-Sponsored I.T. Capstone Program

Giuseppe C. Ng
IST Department, University of Asia and the Pacific
giuseppe.ng@uap.asia

**Abstract**

*Purpose* – The research aims to show the relevance of company client sponsored student projects in the University of Asia and the Pacific Information Technology (UA&P IT) Capstone Program through the use ofan Agile Methodology with Scrum Approach.

*Method* – The modified program is employed on two batches with content analysis and survey results as benchmarks.

*Results* – Surveys at the end of the sprints for both clients and students revealed that the length of the sprint was a critical factor in the development of the information system, and that students learned from addressing additional challenges such as academic load, team pressure and communication issues.

*Conclusion* – Over-all results showed that clients were impressed and keen to adopt the student works.

*Recommendations* – Maintainability aspects of the research can be analyzed for future studies. Increasing the sample size with additional batches could lead to discovery of additional factors not previously seen.

*Research Implications* – The research could help improve other Capstone Programs while improving communication with company clients.

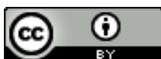

*Keywords* – Agile, Company Client Sponsored Student Projects, Scrum, Sprint, Capstone Program

## INTRODUCTION

The University of Asia and the Pacific Information Technology (UA&P IT) Capstone Program is designed to present students the opportunity to demonstrate their skill in creating IT solutions to address real-world problems. Based on our inquiry with graduates, however, past Capstone products were not utilized by the clients. Furthermore, client company interest and commitment were difficult to achieve and maintain as student outputs were not to their satisfaction (Alzamil, 2005). While companies could be skeptical about student works, citing inexperience as a reason, othersargue that small businesses could benefit from student works by saving on significant resources (Jones & Davey, 2009). An MBA roundtable survey showed that schools have difficulty finding prospective sponsors for their projects and learning programs (Wilbur, 2016), supporting the thesis that clients consider actual projects more important, thus giving less priority to student projects (Marriska, 2015).

Previous studies on company client sponsored projects suggested favorable results in terms of student learning (Parsons & Lepkowska-White, 2009; Sprague & Percy, 2014; Sprague & Hu, 2015). In addition, Schachter and Schwartz (2009) showed that some clients were impressed by student works, highlighting that the approach could work given an appropriate process.

Sommerville (2016) defined software engineering as the process of designing and implementing systems on time and on budget, and identified Agile methodology with Scrum as an example of an iterative approach to developing and delivering systems. This style of development is quickly becoming the trend in the IT industry due to its high success rate and quick delivery of software products (Chawla, 2016; Denning, 2016; Rigby, Sutherland, & Takeuchi, 2016; Linders, 2017; Gross, Hodgett & Ip, 2017). Previous studies using the iterative development approach were completed without the basis of a company client (Coppit, 2006; Stankovic & Tillo, 2009). We want to understand the Agile approach when adopted in a company-client sponsored environment, as a continuation of our previous study on integrating Agile and the Capstone Program (Ng & Venes, 2017).

Our study aims to answer the following questions:
1. How could an Agile methodology adopted into the Capstone Program be successfully implemented?
2. What are the positive and negative effects on students in the Capstone Program?
3. How does Agile methodology improve client engagement?
4. What are the effects of modifying the length of the Sprint to the students'

workload and Capstone Program?
5. What further improvements could be done to the Capstone Program?

**LITERATURE REVIEW**

*Software Engineering*

Sommerville (2016) used the term software engineering to refer to professional software development, focusing on proper discipline and techniques in the construction of major software systems, and specified specific attributes of a professionally built software system: (1) Maintainability or the ability of the software to evolve to business needs, (2) Usability or the ease of using the system, (3) Dependability and Security to determine the reliability and security aspects of the software, (4) Efficiency, and (5) Acceptability, or the software's readiness for use in the environment it was designed for (Sommerville, 2016; Institute of Electrical and Electronics Engineers Computer Society, 2014).

A study by Dyck and Majchzrak (2012) identified several strategies for the different techniques of software engineering, as follows: (1) sequential, iterative, (2) incremental, (3) participatory, or (4) evolutionary.

*Agile Methodology with Scrum*

Agile methodology is a set of values, rather than a process, defined by the Agile Manifesto (Madden, 2017) that focuses on producing working software and highlights close client collaboration ("Manifesto for Agile Development", n.d., para 1).

Sommerville (2016) describes the integration of Agile and Scrum as an incremental approach to development. By having constant feedback from the clients, the software production process would be able to adapt to changes, thereby addressing the weakness of the waterfall model (Sommerville, 2016). Documentation in the Agile methodology with Scrum consists of user stories and acceptance criteria (Apke, 2015).

The employment of the Scrum framework involves different processes (Smith, 2016) covering four distinct events: (1) Sprint Planning; (2) Sprint; (3) Sprint Review; and (4) Sprint Retrospective.

Sprint Planning is the meeting between the developers and the client where everyone sets goals for the development cycle or Sprint. This includes the selection of user stories, system functionalities defined from the end-user's perspective (Smith, 2016; Apke, 2015), and based on feasibility and priority, after which development proceeds (Smith, 2016).

During development, daily Scrums are held. In the Scrums or short meetings, usually held at the start of the day, progress reports are exchanged, and any issues that arose previously could be addressed. Thus, the client could provide feedback and re-prioritize user stories for work (Smith, 2016).

Once the Sprint completes, the team and the client go into another meeting referred to as Sprint Review and Sprint Retrospective. In the Sprint Review, the team and the client discuss the current state of projects and review all the work and accomplishments in the Sprint, and plan for the next Sprint. The Sprint Retrospective is when the team discusses process improvements on the various issues encountered (Smith, 2016).

### *Agile Methodology with Scrum Adopted Capstone Program*

In the design of the Capstone Program, we mapped the Agile methodology with Scrum into the semester. In the initial stage, students are split into groups and select the topic of their proposal. The Capstone Program duration is slightly over four months, administered in the first semester from August to December. This is a continuation of the initial proposal phase from the previous semester, where students select a suitable company client and present to a select panel of faculty members their IT solution to the chosen client's issues, as in Figure 1.

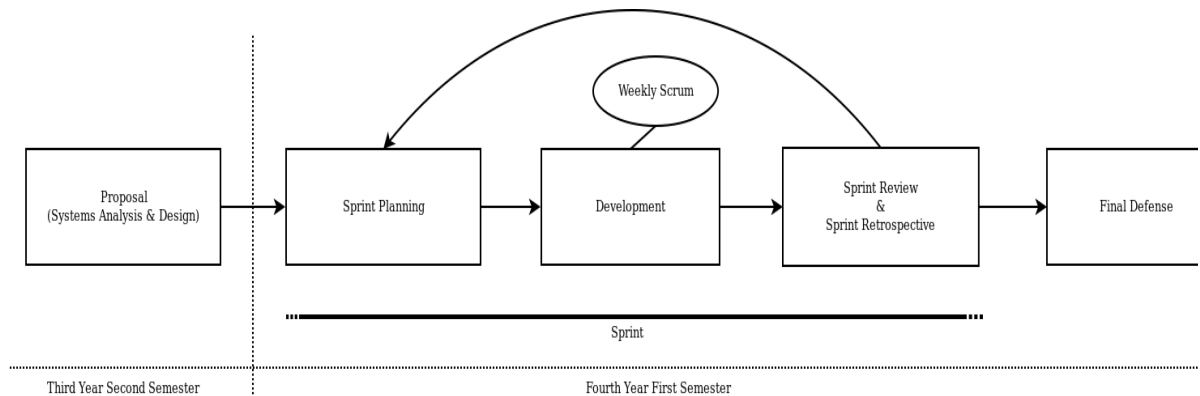

*Figure 1.*   Agile Methodology with Scrum Adopted Capstone Program

Upon acceptance by the panel, the next stage involves the selection of the faculty adviser for the groups as well as the orientation on the Agile methodology with Scrum Adopted Capstone Program. Students perform the initial Sprint Planning. Sprint 1 and the succeeding Sprints occur throughout the whole semester.

The duration of Sprints for the Capstone Program is 20 to 30 working days, with weekends excluded. Short weekly Scrums are held every Wednesday between the students and the faculty adviser to accommodate the former's academic load. At this stage of the Capstone Program, clients are not included in Scrums to avoid demanding too much time from them.

At the end of every Sprint, the student groups create a stable release of their software for Sprint Review. Sprint Review in the Capstone Program is split into two meetings. A client meeting is done first to solicit feedback on the work and re-prioritize user stories for the next Sprint. Clients are also asked to fill an evaluation form based on the student groups' workmanship. In addition, students constantly communicate through additional meetings and electronic communication for clarifications. A second meeting is done with the faculty adviser to cover Sprint Retrospective and Sprint Planning for the next Sprint execution. In total, 15 client meetings and 30 faculty adviser meetings per group on the average was done.

At the end of the development process is the final defense. Usually, this happens alongside the final Sprint as software development typically completes. Student groups are then tasked to attain client acceptance of their work.

For our study, we adopted a post-positivism worldview, because in addition to what is observed, we acknowledge that underlying theories and knowledge of the researcher may influence results (Colin, 2002). In the design and empirical observation of the Capstone Program, Agile methodology and its theories guide the manner in which the program is executed.

We likewise adopted a descriptive research approach to observe and identify the different characteristics of a specific population (Shields & Rangarajan, 2013). Our students would be working with real company clients and experiencing actual development work and all its issues firsthand.

## METHODOLOGY

Before the students could proceed with their Capstone Program, they must first select a suitable company that requires an IT solution. These scenarios are vetted by a panel of faculty members. The following were the criteria of selection:
1. Companies must be locally based or have local representation in the Philippines.
2. Companies must have at least 5 years of operation.
3. Companies must have a problem with a manual business process that can be solved with an IT solution.

At the end of development phase, students were asked to accomplish a survey form as a retrospective of the entire Capstone Program.

The study involved 49 students composed of two batches from Academic Years 2016-2017 (Batch 1) and 2017-2018 (Batch 2), respectively. The following section describes the Capstone Program for the two batches.

### *Batch 1*

There were 31 students split into groups of three or four, making up 10 groups. Development was divided into four Sprints consisting of 20 working days. The list of companies for Batch 1 is covered in our previous work and is shown in Table 1 (Ng & Venes, 2017).

Table 1. Profile of companies (Ng & Venes, 2017)

| Company category | Manual Operation Problems | No. of Years in Operation | Location |
|---|---|---|---|
| Civil works contractor | Training & physical documentation | 10+ | Quezon City |
| Hotel business supplier | Inventory management & client tracking | 40+ | Quezon City |
| Industrial products manufacturer | Handling of HR processes | 10+ | Pasig City (Nationwide) |
| Primary school | Record keeping | 15+ | Pasig City |
| College | Enlistment process | 65+ | Oriental Mindoro |
| College Library | Library Processing | 100+ | Pasig City |
| Preschool Center | Inventory tracking | 5 | Pasig City |
| Dental equipment vendor | Business processes & customer tracking | 35+ | Manila |
| Paper products distributor | Client transaction processing | 30+ | Manila |
| Food manufacturing | Inventory tracking | 35+ | Bulacan |

## Batch 2

There were 18 students split into groups of three, forming 6 groups. Based on the feedback from the previous batch, development was divided into three Sprints consisting of 30 working days. Table 2 shows the list of clients selected.

Table 2. Profile of companies (Batch 2)

| Company category | Manual Operation Problems | No. of Years in Operation | Location |
| --- | --- | --- | --- |
| Auto Supply | Inventory management & client tracking | 20+ | Laguna |
| IT Consultancy | Human resource & payroll | 10+ | Nueva Ecija |
| Rice Distributor | Inventory management, sales tracking, client tracking & delivery tracking | 30+ | Marikina |
| Medicine Distributor | Inventory management, sales tracking & client tracking | 5+ | Quezon City |
| Human Capital Consultancy | Accounting | 15+ | Pasig City |
| Restaurant | Inventory management, point of sales, & sales forecasting | 15+ | Mandaluyong City |

As the feedback on the employment of Sprints and weekly Scrum meetings was positive, the framework for these have not been changed.

For Batch 2, all the groups opted to use a programming language they were adept with (e.g., Java and MySQL). In addition, students were taught how to use git as a source control management tool and some other additional skills as needed. This change was added due to initial feedback from Batch 1.

To determine whether the Agile methodology adoption for our Capstone Program was successful, two aspects must be answered: (1) the company client's impression of the student's project; and (2) the student's acceptance of the Capstone Program scheme. These are established with several evaluation tools.

At the end of each Sprint, client evaluation scores the work based on good software attributes as described by Sommerville (2016), namely: (1) usability, (2) dependability, (3) security, (4) efficiency, and (5) acceptability. We opted to omit maintainability from the evaluation since the criteria cannot be measured within such a narrow time frame. Usability is represented by the criteria 'Easy to learn and use', and 'Appearance is pleasant'. 'Product is Secure' and 'Behavior is appropriate and reliable' are the criteria used for Dependability and Security. The Acceptability metric is measured by 'Complete set of features', 'Included functions are correct', and 'Current build can be deployed'.

The evaluation form used a scale from 1 to 5 for each of the 5 criteria, where 1 is the lowest and 5 is the highest. The scale is described as: (1) very poor, (2) poor, (3) acceptable, (4) good, and (5) very good. 'N/A' score can be given if applicable to the build the students presented. This is converted to 0 for normalization purposes. The acceptable mean score range by the end of a Sprint should be 3.0 to 4.0.

Since company clients may not have sufficient technical background, they may not be able to properly evaluate the criteria. To address possible complexities in the evaluation form from a client perspective, the Batch 2 evaluation form included feedback as to whether the form was easy to understand.

The faculty-in-charge also periodically asked for student insights on the Capstone and the client. These insights are tracked via use of a journal. Along with the survey forms handed to students, these were collated in a table for content analysis. Table 3 illustrates the data analysis plan for the study based on the research questions established.

Table 3. Data Analysis Plan

| Research Questions Nos. | Respon-dents | Data Gathering Instruments | Data Analysis |
| --- | --- | --- | --- |
| (1), (2), (3) | Company Clients | Client evaluation & written feedback Forms | Mean, $t$-test & Content Analysis |
| (2), (3), (4), (5) | Students | Student survey | Content Analysis |

Content analysis was based on categorizing the feedback from the client and students to determine common themes and interesting insights regarding the entire program. Given the notes from meetings with the students and the written evaluation from the client, the comments for both clients and students were collated into a spreadsheet. Each comment was read and categorized and re-categorized based on the emerging themes. Discovering emergent themes is part of the discipline of Complexity Science (Phelan, 2001). This enables us to determine how to improve the program moving forward.

Client feedback is included in truncated form for brevity's sake. From this result, the following research objectives can be achieved: (1) evaluate the client's impression of the student work; (2) identify the positive and negative effects on the student project progress; (3) identify the effects of the Sprint length to the students' output; and, (4) identify changes that can be made to the adopted program.

**RESULTS**

Sprint scores were collated for both batches and the mean score of each group was taken to see the trend of scores and performance. Table 4 and Table 5 present the mean scores of the groups for Batch 1 and Batch 2, respectively. The calculated scores allowed the researcher to determine the trends of the different aspects of the software from the client standpoint.

Table 4. Sprint Mean Scores (Batch 1) (Ng & Venes, 2017)

| Evaluation Criteria | Sprint 1 | Sprint 2 | Sprint 3 | Sprint 4 |
|---|---|---|---|---|
| Easy to learn and use | 4.44 | 4.44 | 4.67 | 4.44 |
| Secure software | 3.22 | 4.00 | 4.56 | 4.11 |
| Behavior is appropriate and reliable | 4.00 | 3.67 | 4.33 | 4.33 |
| Complete set of features | 3.56 | 3.56 | 3.89 | 3.56 |
| Included functions is correct | 3.78 | 3.67 | 3.89 | 4.22 |
| Swift and efficient | 3.89 | 4.00 | 4.11 | 4.22 |
| Appearance is pleasant | 4.44 | 3.89 | 4.22 | 4.22 |
| Current build can be deployed | 3.89 | 3.44 | 3.33 | 3.44 |

Table 5. Sprint Mean Scores (Batch 2)

| Evaluation Criteria | Sprint 1 | Sprint 2 | Sprint 3 |
|---|---|---|---|
| Easy to learn and use | 4.17 | 4.67 | 4.33 |
| Secure software | 4.00 | 4.50 | 4.33 |
| Behavior is appropriate and reliable | 3.50 | 3.50 | 4.00 |
| Complete set of features | 3.33 | 4.17 | 4.50 |
| Included functions is correct | 3.17 | 2.83 | 4.17 |
| Swift and efficient | 3.83 | 3.67 | 4.17 |
| Appearance is pleasant | 4.00 | 4.50 | 4.33 |
| Current build can be deployed | 3.33 | 4.00 | 4.33 |

Figure 2 illustrates the mean score based on Batch 1 while Figure 3 shows the results for Batch 2. One group in Batch 1 was not included in the results due to extraordinary circumstances. This is covered in our previous work (Ng & Venes, 2017). Results and the comparison of the two academic years were based on the software attributes that were covered in the form.

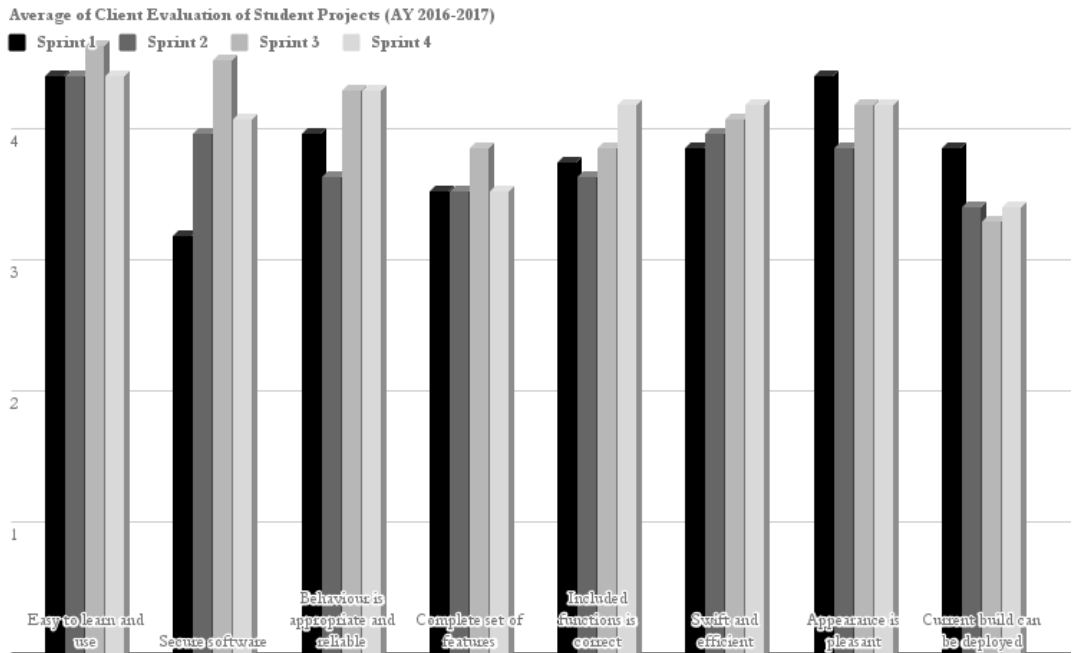

*Figure 2.* Average of Client Evaluation for Student Projects (Batch 1) (Ng & Venes, 2017)

## *Usability*

This software attribute is covered by the criteria: 'Easy to learn and use', and 'Appearance is pleasant'. From Tables 3 and 4, the mean score feedback from the client is above 4.0. There was one instance in Sprint 2 Batch 1 where the mean score dipped below 4.0. The reason was that some groups decided to focus more on functionality than interface work, resulting in major interface changes due to additional tasks and possible misinterpretation of the requirements.

With the consistency of the rest of the scores, it could be noted that the clients were impressed with the front-end work. Nonetheless, there is a slight variance with the scores to suggest that improvements were made through the Sprints. A t-test showed that the variance presented no significant difference between the two batches.

As proposed by Apke (2015), client feedback is integral to the success of the Agile methodology. With frequent client meetings, more suggestions were reflected, showing a slight increase in the scores. With the last Sprint, there was a slight dip in both academic years. This was due to the user interface having no more notable or significant changes, leaving a less impactful impression on the clients.

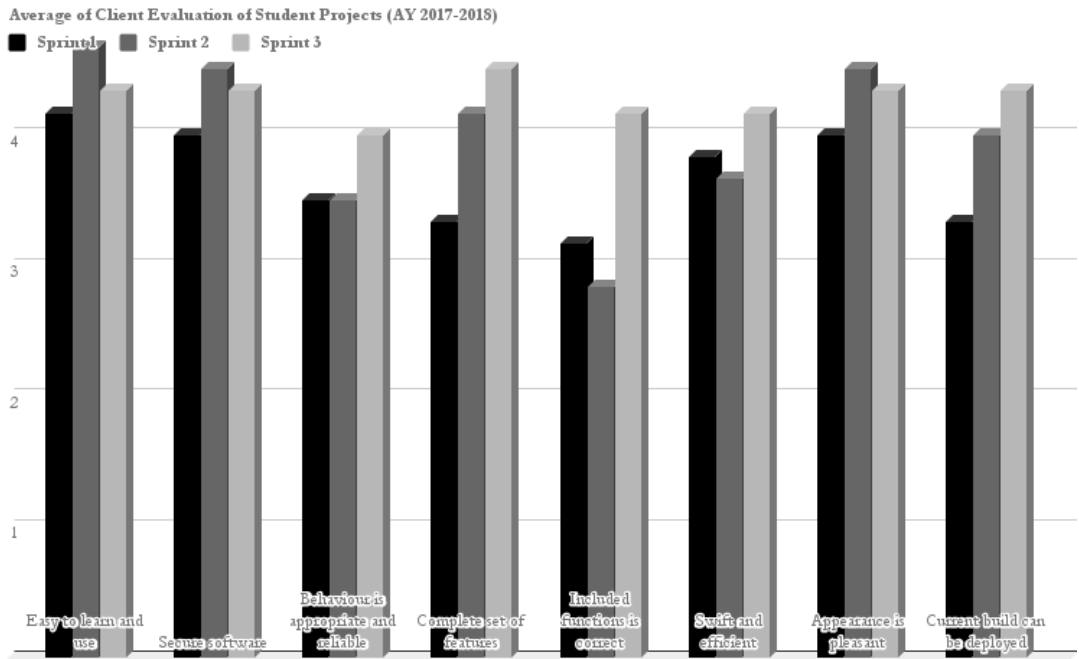

*Figure 3.*   Average of Client Evaluation for Student Projects (Batch 2)

## *Dependability and Security*

This software attribute is covered under the scores for the criteria 'Product is Secure' and the 'Behavior is appropriate and reliable'. From Figures 3 and 4, the scores showed a comparably upward trend. Not all major security features had been in place in earlier Sprints, but these aspects were improved with further Sprints. Also, with every Sprint, more bugs were being addressed, leading to a more reliable product, thus improving client confidence.

As explained by Sommerville (2016), a software product is incrementally built over each Sprint. The development progress of the software of the student Capstone reflected this.

## *Efficiency*

The mean scores from both batches showed general improvement from the first to the last Sprint. As explained in previous sections, the evolution of the software meant general improvements to the product in the eyes of the client.

There was, however, a dip in the Sprint 2 score for Batch 2 which came from two groups whose clients requested removal of several functionalities and changes in the process and the account permission systems. This change from the client reflected a shift in demands and circumstances that, as Sommerville (2016) pointed out, is characteristic of customer involvement.

*Acceptability*

This software attribute is dependent on the following criteria: 'Complete set of features', 'Included functions are correct', and 'Current build can be deployed'.

In the 'Included functions are correct' on Table 4, there is a significant dip in the score on Sprint 2 which was caused by the evaluation of a different stakeholder. According to the students, their contact in the company was not available and so when they met with a different stakeholder in the same company, the person provided a different perspective of the system. The score reflected that not all the insights provided by the previous client contact was not to his satisfaction.

On Table 3, the trend of the scores for both 'Complete set of features' and 'Current build can be deployed' were in the range of 3 whereas in Table 4, the scores increased all the way past 4. In the 'Complete set of features' criteria, one group consistently got 'N/A' scores from the client due to unfinished features. Over time, the client had requested different features, which affected the evaluation metrics the client had for the groups. Batch 1 students often complained that the 20 working days window was too tight, and this seemed to influence how much work they could accomplish in between Sprints. With every Sprint, the client would bring additional stakeholders that would generate new feature requests and alter the scope and evaluation metrics. Using a t-test, these criteria showed significant difference between the two batches, as we further explain below.

Sommerville (2016) proposed that tasks are scheduled for each Sprint. Whereas Batch 1 had a severely limited time window to address this, the Batch 2 program was structured to allow 30 working days per Sprint. This meant more work could be scheduled in the Sprints and this led to more complete software during end of Sprint meetings with the client.

## DISCUSSION

*Company Client Perspective*

Table 6 illustrates the client feedback for Batch 1 and Table 7 shows the client feedback for Batch 2. Content analysis was performed to determine the nature of feedback from the clients. These were categorized into: (1) Project Comments; (2) Easy to use; (3) Deployment Sentiments; (4) Front end issues; (5) Functionality issues; (6) New feature requests. Feedback written here have been trimmed for brevity.

Table 6. Company Client Feedback (Batch 1) (Ng & Venes, 2017)

| Criteria | Client Responses |
|---|---|
| (1) Project Comments | Excited to use the system; Overall, very good! I look forward to more improvements; I think the product will help us. |
| (2) Easy to use | Easy to use, User friendly |
| (3) Deployment Sentiments | Product is exactly the same as the demands of the Company; Comprehensive; Recommendable for future use |
| (4) Front-end issues | Front end needs improvement; Change color scheme Add website footer; Change product layout |
| (5) Functionality issues | Needs improvement in shopping cart; User information must be editable; Specific section is still lacking features |
| (6) New feature requests | Add report with date filtering; Add email functionality for sales; Search applicant; Add support for archive data |

Table 7. Company Client Feedback (Batch 2)

| Criteria | Client Response |
|---|---|
| (1) Project Comments | Students are professionals and competent; The system is good but not complete; Great work! |
| (2) Easy to use | The user can easily navigate the functionality of the program. |
| (3) Deployment Sentiments | Excited to use the system in the company; We plan to deploy in January 2018; We look forward to deploying the system. |
| (4) Front end Issues | Fix user interface to make the system more user friendly; Update the CSS of the system; Add navigation bar. |
| (5) Functionality Issues | Test all functionalities before the client meeting; Fix permission issues; Add error handling; Prioritize generation of sales report; Remove archiving of products. |
| (6) New feature requests | Add viewing of products; Add notification of stocks; Add PDF report for printing; Add different user accounts. |

From the written feedback, the clients' overall sentiments covered two general aspects. The first one is software related. As could be seen in Tables 5 and 6, criteria (2), (4), (5) and (6) pertain to software improvements and suggestions from the client. User interface comments and suggestions, (2) and (4) respectively, were extensively considered as these were visible components of the software.

The second aspect pertains to the accuracy of the work and the project's overall accomplishments by both batches. Clients expressed great interest in the deployment of the system as it addressed the problems identified, as could be seen in criteria (3). From criteria (1) and (4), it was evident that clients were keen to adopt the software produced by the students. This is further established by the mean scores garnered in the Sprints.

The repeated Sprint Review meetings also created a better working relationship between the clients and the students. These results are consistent with Smith (2016) who asserted that Agile's success is from customer feedback and acceptance, and Apke (2015), who argued that satisfying the customer and gaining full trust are of great importance to success.

*Student Feedback and Analysis*

Considerable student feedback were collected from survey and faculty observation throughout the development of the software. The results are shown in Table 8.

Table 8. Student Observation of Clients

| Criteria | Client Response |
| --- | --- |
| Client disinterest | Client seemed unwilling to participate perhaps because of additional work for them; Client seemed to think the project was unimportant. |
| Incorrect Project Impressions | Client had no idea of the system being built; Client was already expecting substantial work; On first Sprint meeting, the client thought the system was already done. |
| Positive Client Interactions | Clients were excited to work with us; Client interactions were smooth and very responsive. |

Meeting clients was challenging to students as the client contact person was at times not available. Constant communication with their respective clients, however, resulted in a marked increase of client interest in the student's projects. Additional suggestions by the client improved the development direction of the system.

The student groups in Batch 1 opted for various programming languages and frameworks with which they were not familiar, thus causing training overhead. For Batch 2, since all groups utilized tools familiar to them, the learning curve focused on learning additional tools for more complex functionalities.

Time management was an ongoing challenge as students had to manage their other academic loads while maintaining project development progress. Batch 1 students suggested extending the Sprint because the 20-working day window was simply too tight. This was adjusted for the next academic year to 30 working days per Sprint. While students still cited academic load as challenging, they had more time to work and manage their Sprints.

The varying skill levels of members compounded development progress as not all members could produce suitable outputs in a timely fashion. This led to disproportionate distribution of work, delays, and burnout, causing internal conflicts in both Batches. With Batch 2, the longer 30-day duration of Sprints allowed some group members to slacken and take on fewer tasks. Some of these issues were not raised during Scrum meetings and Sprint Retrospectives, causing frequent delays and additional stress.

One group in Batch 2 identified this weakness early and resorted to building abstraction layers of code to make development work easier for the rest of the members. Other groups mentioned that this practice could be emphasized for future Capstone Programs.

Communication among members was a challenge for some, especially when there is difficulty finding a common working schedule. The use of social media, while successful for some groups, proved to be ineffective for others. In Batch 2, Slack was adopted as an additional communication tool that some appreciated, but which others felt unnecessary.

Feedback regarding the employment of Agile methodology and Sprints was positive. The weekly Scrum meetings allowed students to keep abreast of developments and other ongoing issues. The Sprints themselves put the discipline of deadlines in place and allowed students to focus on maintaining the progress of their development. Students were overwhelmingly unanimous in proposing that Agile methodology be adopted for the next Capstone Program.

## *Project Issues and Failure in Batch 1*

One of the groups in Batch 1 resulted in project failure. Due to the exacerbation of development troubles, they were not able to produce satisfactory builds leading to the client dropping their support for the project. This was caused by multiple factors.

The most significant factor for the failure was the complete violation of the Agile manifesto. The manifesto establishes that the team should focus on individuals, interactions and working software over the tools and the plan ("Manifesto for Agile Development", n.d., para 1). This group had instead devoted much of their time on selecting which language to use and switching technologies too frequently, thereby discarding much of their previous work.

Due to the lack of focus on individuals and interactions, communication became a significant hurdle. The team could not agree on common working times, and the use of social media and asynchronous messaging proved ineffective. Most of the messages left were only partially read, if seen at all. They also developed personal gripes against each other that built over time and without any way of resolving conflicts. Even though weekly Scrums were held, none of these internal conflicts were raised and brought out in the open, hence getting worse over time.

The development woes led to the group missing Sprints and, eventually, to cease communicating with the client. Supporting the argument of Apke (2015) about the importance of client feedback in Agile methodology, this eventually led to the project's failure as it no longer reflected the client company's changing needs and circumstances.

## *Improvement from Batch 1 to Batch 2*

From the comparison of the mean scores of the two batches, there was evidence showing that Batch 2 performed better than Batch 1. Table 9 shows the *t*-test results, formatted for brevity's sake.

Table 9. *t*-test of Batch 1 and Batch 2 software criteria

| Software Criteria | Sprint 1 | | | Last Sprint | | |
|---|---|---|---|---|---|---|
| | t Stat | P (T<=t) one-tail | P (T<=t) two-tail | t Stat | P (T<=t) one-tail | P (T<=t) two-tail |
| Ease of Use | 1 | 0.181 | 0.363 | 0.542 | 0.305 | 0.610 |
| Secure Software | -1.274 | 0.129 | 0.258 | -0.542 | 0.305 | 0.610 |
| Behavior is appropriate and reliable | 1.348 | 0.117 | 0.235 | 1.463 | 0.101 | 0.203 |
| Complete Set of Features | 0.237 | 0.411 | 0.822 | -1.659 | 0.078 | 0.157 |
| Included functions are correct | 0.790 | 0.232 | 0.465 | 0 | 0.5 | 1 |
| Swift and Efficient | -0.191 | 0.428 | 0.856 | 0 | 0.5 | 1 |
| Appearance is pleasant | 1 | 0.181 | 0.363 | 0 | 0.5 | 1 |
| Current build can be deployed | 0.590 | 0.290 | 0.580 | -2.390 | 0.031 | 0.062 |

From the *t*-tests conducted, there were only two software criteria that showed significant difference: "Complete Set of Features" and "Current build can be deployed". Both evaluation criteria fall under the software attribute "Acceptability". The analysis of the last Sprint showed that clients were far more convinced with the software produced by Batch 2 than those in Batch 1.

Much of the improvement could be attributed to two factors: (1) the increase of working days in the Sprint from 20 to 30 working days, which allowed for more leeway for the students to manage their academic loads and development loads and adjust to unforeseen circumstances; and (2) the reduction of the student's learning curve in adopting tools and technical skills, which was addressed in two ways. First, students began development using familiar languages and development platforms. Second, additional training was provided for platforms that had to be learned such as git.

## CONCLUSIONS AND RECOMMENDATIONS

Based on the findings and results, the company clients for both Batch 1 and Batch 2 have been very impressed with the outputs of our students. Student feedback was overwhelmingly positive about the usefulness of Agile development. The positive feedback from clients and students proved that the adopted Agile methodology Capstone Program is successful.

Using Sprints, clients saw the evolution of the software being produced, thus increasing their interest and involvement in the students' project, and resulting in over-all positive client evaluations. It could be concluded, therefore, that the Agile methodology did improve client engagement.

From the students' standpoint, many realized the effectiveness of having a planned way of software development. Employing the Agile methodology of software development, the student groups were able to keep track and maintain progress and development. They were exposed to several aspects of development issues, from team members' dynamics to software complexities, and learned how to adapt to the situations and learn new skills when needed. In fact, one of the clients showed keen interest engaging the students for further work.

The overhead of learning new languages and tools that existed in Batch 1 was addressed in Batch 2 with the use of known programming languages. Without the need to learn new languages, the students were able to work directly on producing the system modules. This allowed them to focus on the learning and exploration of specific tools needed to complete the system.

As to the negative effects of the program on the students, the Agile methodology demanded a lot of time and resources from each student, making managing academic loads a challenge. This was partially addressed with the extension of the Sprint length from 20 working days to 30 working days. There were no complaints regarding the Sprints being too short in Batch 2, whereas student feedback in Batch 1 showed that 20 working days were simply too short a period for them. Having ten additional days allowed students to better manage their academic workload. However, there were complaints about the Sprint being too long, thus encouraging some group members not to work as hard.

Among the problems that surfaced among the students, the most significant could be the refusal or inability to bring up technical issues and team issues, leading to additional stress and burnout. Even though Scrum and Sprint Review meetings were held, some groups failed to take advantage of these team management opportunities to address their team issues. The issue of transparency in communication would have to be emphasized in future Capstone Programs.

In summary, we conclude the following:
1. Adoption of an Agile methodology in the Capstone Program is not only possible but also yields successful results.
2. The study identified several positive effects from the good quality software produced to several valuable learning aspects of the Capstone Program, and also some issues that need to be addressed, particularly on time management and team management skills, especially communication transparency.
3. Employment of Agile Methodology improved client interest and involvement.
4. Extension of the Sprint to 30 working days had positive results from the standpoint of managing student academic workloads and project software development, though it may have inadvertently influenced imbalanced work distribution.

A final set of recommendations is for this research to be extended to cover maintainability aspects, and to include more batches in the study to allow further refinement of the program. Also, further studies can find the correlation between the number of meetings and evaluation ratings.

## IMPLICATIONS

The research helped demonstrate that it is possible to create a Capstone Program that integrates company interactions with student projects that generate value for both sides. By laying out the framework for the Capstone Program, other institutes may examine and adopt practices applicable in their case.

## ACKNOWLEDGEMENT

We would like to thank the participating BSIT senior students, the companies that supported the Capstone Program, Dr. Ligot, and the Department of Information Science and Technology of the University of Asia and the Pacific for supporting this research.